\documentclass[12pt]{article}
\usepackage{amsmath,caption,centernot}
\usepackage[top=1.35in, bottom=1.35in, left=1.35in, right=1.35in]{geometry}

\DeclareCaptionStyle{italic}[justification=centering]
 {labelfont={bf},textfont={it},labelsep=colon}
\usepackage{graphicx,psfrag,epsf}
\usepackage{enumerate}
\usepackage{natbib}
\usepackage{url} 
\usepackage{booktabs, subfig, bm, paralist,mathpazo,tikz,todonotes,longtable,microtype} 
\usepackage[pdftex,colorlinks=true]{hyperref}
\definecolor{darkblue}{rgb}{0,0,.6}
\hypersetup{citecolor=darkblue,linkcolor=darkblue,urlcolor=darkblue}

\newcommand{\blind}{0}

\graphicspath{{plots/}}

\newsavebox\CBox
\def\textBF#1{\sbox\CBox{#1}\resizebox{\wd\CBox}{\ht\CBox}{\textbf{#1}}}

\definecolor{a0}{rgb}{0.0, 0.5, 0.0}
\definecolor{bistre}{rgb}{0.24, 0.17, 0.12}
\definecolor{amethyst}{rgb}{0.6, 0.4, 0.8}
\definecolor{blue-violet}{rgb}{0.54, 0.17, 0.89}
\definecolor{Rcolor}{RGB}{150,160,190}
\definecolor{blush}{rgb}{0.87, 0.36, 0.51}
\definecolor{brightturquoise}{rgb}{0.03, 0.91, 0.87}
\definecolor{burntorange}{rgb}{0.8, 0.33, 0.0}
\begin{document}

\def\spacingset#1{\renewcommand{\baselinestretch}%
{#1}\small\normalsize} \spacingset{1}

\if0\blind
{
  \title{\bf Bootstrap methods for stationary functional time series}
  \author{Han Lin Shang\footnote{Postal address: Research School of Finance, Actuarial Studies and Statistics, Australian National University, Canberra ACT 2601, Australia; Telephone number: +61-2-612 50535; Fax number: +61-2-612 50087; Email: hanlin.shang@anu.edu.au} \\
    Research School of Finance, Actuarial Studies and Statistics \\
    Australian National University}
  \maketitle
} \fi

\if1\blind
{
  \bigskip
  \bigskip
  \bigskip
  \begin{center}
    {\LARGE\bf Title}
\end{center}
  \medskip
} \fi

\bigskip
\begin{abstract}
Bootstrap methods for estimating the long-run covariance of stationary functional time series are considered. We introduce a versatile bootstrap method that relies on functional principal component analysis, where principal component scores can be bootstrapped by maximum entropy. Two other bootstrap methods resample error functions, after the dependence structure being modeled linearly by a sieve method or nonlinearly by a functional kernel regression. Through a series of Monte-Carlo simulation, we evaluate and compare the finite-sample performances of these three bootstrap methods for estimating the long-run covariance in a functional time series. Using the intraday particulate matter (PM$_{10}$) data set in Graz, the proposed bootstrap methods provide a way of constructing the distribution of estimated long-run covariance for functional time series.

\vspace{.2in}
\end{abstract}

\newpage
\spacingset{1.45} 

\section{Introduction}
\label{intro}

Functional data often arise from measurements obtained by separating an almost continuous time record into natural consecutive intervals, for example days, weeks or years; see \cite{HK12}. The functions thus obtained form a time series $\{\mathcal{X}_i,i\in Z\}$ where each $\mathcal{X}_i$ is a random function $\mathcal{X}_i(t)$ for $t$ lies within a function support range $\mathcal{I}$. We refer to such data structures as functional time series, examples of which include daily price curves of a financial stock in \cite{KZ12}, monthly sea surface temperature in climatology in \cite{SH11}, and yearly age-specific mortality rates in \cite{HU07}. 

A central issue in the analysis of such data is to take into account the temporal dependence of the functional observations denoted by $\bm{\mathcal{X}} = (\mathcal{X}_1, \dots, \mathcal{X}_n)^{\top}$. Due to this temporal dependence, even the most elementary statistics become inaccurate; an example is estimating the unknown mean function $\mu = \text{E}(\mathcal{X})$ of a functional time series. The sample mean $\bar{\mathcal{X}}_n = \frac{1}{n}\sum^n_{i=1}\mathcal{X}_i$ is not the most efficient estimator in the presence of strong serial correlation. As shown in \cite{Politis03}, one can decompose a functional time series as $\mathcal{X}_i = \mu + \varepsilon_i$. When $\varepsilon_i$ are independent and identically distributed (IID), $\bar{\mathcal{X}}_n$ is the ordinary least squares estimator of $\mu$. However, when $(\varepsilon_1,\dots,\varepsilon_n)$ are correlated, the best linear unbiased estimator of $\mu$ is the generalized least squares estimator. Let $\bm{I} = (1,\dots,1)^{\top}$ and $\bm{\Gamma}_n$ denotes the (unknown) covariance operator of a time series of functions $\bm{\mathcal{X}}$ with $i, j$ element given by $\gamma(i-j) = \text{Cov}(\mathcal{X}_i, \mathcal{X}_j)$. The generalized least squares estimator is then given by $\widehat{\mu} = \left(\bm{I}^{\top}\bm{\Gamma}_n^{-1}\bm{I}\right)^{-1}\bm{I}^{\top}\bm{\Gamma}_n^{-1}\bm{\mathcal{X}}$, where $^{\top}$ symbolizes matrix transpose. It is obvious that $\widehat{\mu}$ is a weighted average of $\bm{\mathcal{X}}$, where the weights depend on the unknown covariance operator. As many statistics require the correct calculation of sample mean function, the serial correlation in functional time series poses challenges in the calculations of other summary statistics, such as long-run covariance function considered in Section~\ref{sec:2}.

In functional time series, a main task is to make inference about the sampling distribution of $\widehat{\vartheta}_n$, a statistic estimating the parameter of interest $\vartheta$. Not only it is important to obtain a consistent estimator, we are also interested in estimating the variability associated with $\widehat{\vartheta}$ and constructing its confidence intervals (CIs) or carrying out a hypothesis test (see, for example, \cite{BHK09}, \cite{HH06}, \cite{HKR14}, \cite{PS15}). When such a problem arises, resampling methodology especially bootstrapping turns out to be the only practical alternative (see, for example, \cite{CFF06}, \cite{GGC13}, \cite{MP11}, \cite{Shang15}) by using sampling from the sample to model sampling from the population. However, the dependence in functional time series complicates matters because the bootstrap must be modified so that the resampled functional time series has an asymptotically similar dependence structure to the original functional time series. 

Bootstrapping functional data is not new, but it has been studied mostly on independent functional data (see e.g., \cite{BHK09}, \cite{CFF06}, \cite{PS15}, \cite{Shang15}). There are relatively fewer studies on bootstrap techniques for functional time series, noticeably the work of \cite{RAV+16} and \cite{RAV15} who extended the stationary bootstrap of \cite{PR94} to functional time series. The essential idea of stationary bootstrap is that the resampled functional time series contains arbitrary length blocks of consecutive observations from the original functional time series, where the average length of the blocks grows as sample size becomes large. However, it relies on stationarity condition and it can be difficult to determine the optimal length of blocks, which is a form of smoothing parameter. Using too small block length will corrupt the dependence structure, increasing the bias of the bootstrap method, choosing block length too large will result a method which has relatively high variance and consequent inaccuracy; see \cite{PRW99}. 

Bootstrapping functional time series is receiving increasing attention in functional data literature, as shown by three concurrent working papers. \cite{FN16} proposed a residual-based bootstrap for functional autoregressions and showed that the empirical distribution of the centered sample innovations converges to the distribution of the innovations with respect to the Mallows metric. \cite{Paparoditis16} also considered the functional autoregressions and derived bootstrap consistency as the sample size and order of functional autoregression both tend to infinity. From a nonparametric viewpoint, \cite{ZP16} proposed a kernel estimation of first-order nonparametric functional autoregression model and its bootstrap approximation. 

To contribute to this field, this article provides an informative account of three techniques from a computational perspective that merge the ideas of functional time series analysis and bootstrap techniques for univariate time series in Section~\ref{sec:3}. Illustrated by simulation studies in Section~\ref{sec:4}, we evaluate and compare the finite-sample performances between these three bootstrap techniques. In Section~\ref{sec:5}, we apply these bootstrap techniques to an intraday PM$_{10}$ data set in Graz, Austria. Conclusions are given in Section~\ref{sec:6}, along with some ideas on how the methodology presented here can be further extended in this exciting area of research.

\section{Estimation of long-run covariance}\label{sec:2}

\subsection{Notation}

It is commonly assumed that random functions are sampled from a second-order stochastic process $\mathcal{X}$ in $L^2$, where $L^2$ is the Hilbert space of square-integrable functions. Each realization $\mathcal{X}_i$ satisfies the condition $\|\mathcal{X}_i\|^2 = \int_{\mathcal{I}} \mathcal{X}^2_i(t) dt < \infty$ with a function support range $\mathcal{I}$, inner product $\langle f,g\rangle=\int_{\mathcal{I}}f(t)g(t)dt$ for any two functions, $f$ and $g\in L^2(\mathcal{I})$ and induced squared norm $\|\cdot\|=\langle \cdot,\cdot\rangle$. All random functions are defined on a common probability space $(\Omega, A, P)$. The notation $\mathcal{X}\in L_H^p(\Omega, A, P)$ is used to indicate that for some $p>0$, the condition $\text{E}(\|\mathcal{X}\|^p)<\infty$. When $p=1$, $\mathcal{X}$ has the mean curve $\mu = \text{E}(\mathcal{X})$; when $p=2$, $\mathcal{X}$ has the covariance operator $\mathcal{K}(s,t) = \text{Cov}[\mathcal{X}(s),\mathcal{X}(t)] = \text{E}\{[\mathcal{X}(s) - \mu(s)][\mathcal{X}(t) - \mu(t)]\}$, where $s, t\in \mathcal{I}$.

\subsection{Long-run covariance estimation}\label{sec:2.2}

In order to provide a formal definition of the long-run covariance function, suppose that $\{\mathcal{X}_i(t), t\in \mathcal{I}\}_{i\in Z}$ is a set of stationary and ergodic functional time series. The long-run covariance function is defined as
\begin{align*}
\mathcal{K}(s,t) &= \sum^{\infty}_{\ell = -\infty} \gamma_{\ell}(u,s), \\
\gamma_{\ell}(u,s) &= \text{cov}(\mathcal{X}_0(u), \mathcal{X}_{\ell}(s)),
\end{align*}
and is a well-defined element of $L^2$ under mild weak dependence and moment conditions. Via right integration, $C$ defines a Hilbert-Schmidt integral operator on $L^2$ given by
\begin{align}
C(f)(u) = \int \mathcal{K}(s,t) f(s)ds, \label{eq:C_opt}
\end{align}
whose eigenvalues and eigenfunctions are related to the dynamic functional principal components defined in \cite{HKH15}, and~\eqref{eq:C_opt} provides asymptotically optimal finite dimensional representations of the sample mean of dependent functional data.

It is of interest in practice to estimate $C$ from a finite sample $\mathcal{X}_1, \dots, \mathcal{X}_n$. Given its definition as a bi-infinite sum, a natural estimator of $C$ is
\begin{align*}
\widehat{\mathcal{K}}_{h,q}(s,t) = \sum^{n-1}_{\ell = -(n-1)} W_q\left(\frac{\ell}{h}\right)\widehat{\gamma}_{\ell}(u,s), 
\end{align*}
where $h$ is called the bandwidth parameter,
\[ \hspace{-.3in}  \widehat{\gamma}_{\ell}(u,s) = \left\{ \begin{array}{ll}
         \frac{1}{n}\displaystyle\sum_{j=1}^{n-\ell}\left[\mathcal{X}_j(u) - \overline{\mathcal{X}}(u)\right]\left[\mathcal{X}_{j+\ell}(s) - \overline{X}(s)\right], & \mbox{$\ell \geq 0$};\\
         \frac{1}{n}\displaystyle\sum_{j=1-\ell}^n \left[\mathcal{X}_j(u) - \overline{X}(u)\right]\left[\mathcal{X}_{j+\ell}(s) - \overline{\mathcal{X}}(s)\right], & \mbox{$\ell < 0$}.\end{array} \right. \] 
is an estimator of $\gamma_{\ell}(u,s)$, and $W_q$ is a symmetric weight function with bounded support of order $q$. Mild conditions must be assumed on the bandwidth parameter $h$ in order for $\widehat{\mathcal{K}}_{h,q}$ to be a consistent estimator of $C$ in norm, that is $h=h(n) \rightarrow \infty$ as $n\rightarrow \infty$ and $h(n) = o(n)$. However, its choice can greatly affect the performance of the estimator in finite samples, and hence \cite{RS16} developed a two-step data-driven approach to select $h$ in order to minimize the estimation error, and showed its asymptotic consistency. In the first step, a pilot bandwidth is estimated using the flat-top kernel of \cite{PR96} as the initial weight function and $h=n^{\frac{1}{5}}$ as the initial bandwidth. In the second step, the first-order Bartlett kernel is chosen as the optimal final weight function.

\section{Some resampling techniques}\label{sec:3}

\subsection{Bootstrap functional principal component scores}\label{sec:me}

Among many techniques for modeling the variability of functional time series, model based on Karhunen-Lo\'{e}ve decomposition are commonly used (see \cite{HKH15} among others). A Karhunen-Lo\'{e}ve expansion of $\mathcal{X}$ evaluated at $t\in \mathcal{I}$ is expressed by
\begin{align}
\mathcal{X}(t) &= \mu(t) + \varepsilon(t) \notag \\
	&=  \mu(t) + \sum^{\infty}_{j=1}\beta_j\phi_j(t), \label{eq:1}
\end{align}
with the mean function $\mu(t) = \text{E}[\mathcal{X}(t)]$ and the basis functions $(\phi_1(t),\phi_2(t),\dots)$ are the orthonormal eigenfunctions of the long-run covariance kernel $\mathcal{K}(s,t)$. The long-run covariance kernel $\mathcal{K}(s,t)$ can be expressed as
\begin{align*}
\int_{\mathcal{I}} \mathcal{K}(s,t)\phi_j(t)dt &= \lambda_j\phi_j(s), \\
\mathcal{K}(s,t) &=\sum^{\infty}_{j=1}\lambda_j\phi_j(t)\phi_j(s),
\end{align*}
where $\lambda_j\geq 0$ is a set of eigenvalues in a decreasing order, and the condition $\int_{\mathcal{I}}\text{E}\left[\mathcal{X}^2(t)\right]dt <\infty$ entails $\sum^{\infty}_{j=1}\lambda_j <\infty$. The principal component scores $\beta_j$ in~\eqref{eq:1} is given by the projection of $\mathcal{X}-\mu$ in the direction of the $j^{\text{th}}$ eigenfunction $\phi_j$, i.e., $\beta_j=\langle \mathcal{X}-\mu, \phi_j\rangle$. The principal component scores $(\beta_1,\beta_2,\dots)$ consist of uncorrelated sequences of random variables with zero mean and finite variance. 

In practice, we often observe a time series of functions at regular and dense grid in \cite{RS05} or irregular and sparse grid in \cite{YMW05}. In the case of irregular and sparse grid, one can implement a nonparametric smoothing technique, such as local linear smoother, to obtain a regular and dense grid (see for example, \cite{YMW05}). Due to simplicity, we concentrate on the regular and dense grid, where the number of grid points is often larger than the sample size. A time series of functions $\bm{\mathcal{X}}(t) = \left\{\mathcal{X}_1(t),\dots,\mathcal{X}_n(t)\right\}$ can be decomposed as
\begin{equation*}
\mathcal{X}_i(t) = \widehat{\mu}(t) + \sum^{n}_{k=1}\widehat{\beta}_{i,k}\widehat{\phi}_k(t), \qquad i=1,\dots,n,
\end{equation*}
where $\widehat{\mu}(t) = \frac{1}{n}\sum^n_{i=1}\mathcal{X}_i(t)$, $\widehat{\bm{\beta}}_k = \big(\widehat{\beta}_{1,k},\dots,\widehat{\beta}_{n,k}\big)$ are estimated from empirical covariance function and $(\widehat{\bm{\beta}}_1,\dots,\widehat{\bm{\beta}}_K)$ are uncorrelated series, and $\big\{\widehat{\phi}_1(t),\widehat{\phi}_2(t),\dots,\big\}$ represents a set of estimated functional principal components. Since principal component scores are considered as surrogates of original functional time series, these principal component scores capture the dependence structure inherited in the original functional time series (see also \cite{Paparoditis16} and \cite{NG15}). By adequately bootstrapping these principal component scores, we can generate a set of bootstrapped functional time series, conditional on the estimated mean function and estimated functional principal components from the observed functional time series.

Among many bootstrap techniques for time series of principal component scores, we implement a maximum entropy (ME) bootstrap proposed by \cite{Vinod04}. The advantages of the ME bootstrap for univariate time series are:
\begin{enumerate}
\item[(i)] stationarity condition is not required;
\item[(ii)] bootstrap technique computes ranks of a time series, thus it is robust against outliers of the principal component scores;
\item[(iii)] bootstrap samples satisfy the ergodic theorem, central limit theorem and mean preserving constraint;
\item[(iv)] bootstrap samples are adjusted so that the population variance of the ME density equals that of the original data.
\end{enumerate}

For each series of the estimated principal component scores, we approximate its probability distribution, as a measure of uncertainty of a random variable. In the information theory literature, the measure of uncertainty is known as entropy in \cite{Shannon48}. Instead of assuming a parametric distribution for the principal component scores, it is more objective to choose the functional form of this probability distribution, which maximizes the Shannon entropy subject to mass and mean-preserving constraints. This is the idea of ME.

In order to use the ME to construct a bootstrap method valid for time series with any level of persistence, we must take into account data information and preserve persistence. \cite{VL09} introduce a ME bootstrap method to address both points. The algorithm can be summarized below.
\begin{enumerate}
\item[1)] Sort each set of the estimated principal component scores in increasing order to create order statistics $x_{(i)}$ and store the ordering index vector.
\item[2)] Compute intermediate points $z_t = \frac{x_{(i)}+x_{(i+1)}}{2}$ for $i=1,\dots,n-1$ from the order statistics.
\item[3)] Compute the trimmed mean, denoted by $m_{\text{trim}}$ of deviations $x_i - x_{i-1}$ among our consecutive observations. Compute the lower limit for left tail as $z_0 = x_{(1)} - m_{\text{trim}}$ and upper limit for right tail as $z_n = x_{(n)}+m_{\text{trim}}$. These limits become the limiting intermediate points.
\item[4)] Compute the mean of the ME density within each interval such that the ``mean-preserving constraint" is satisfied. Interval means are denoted as $\{m_1, \dots, m_n\}$. The means for the first and last intervals have simpler formulas:
\[ \left\{ \begin{array}{l}
         m_1=\frac{3}{4} x_{(1)}+\frac{1}{4} x_{(2)} \\
        m_{\eta} = \frac{1}{4}x_{(\eta-1)} + \frac{1}{2} x_{(\eta)} + \frac{1}{4} x_{(\eta+1)},  \ \eta=2,\dots,n-1, \\
        m_n = \frac{1}{4} x_{(n-1)} + \frac{3}{4} x_{(n)}
        \end{array} \right. \] 
\item[5)] Generate random numbers from Uniform$[0,1]$, compute sample quantiles of the ME density at those points and sort them.
\item[6)] Re-order the sorted sample quantiles by using the ordering index of step 1). This recovers the temporal dependence relationships of the originally observed data.
\item[7)] Adjust the variance of bootstrap samples so that the population variance of the ME density equals that of original data, see \cite{Vinod13}.
\item[8)] Repeat steps 2) to 7) several times.
\end{enumerate}  

Computationally, the \verb meboot.pdata.frame \ function in the \textit{meboot} package in \textsf{R} \cite{Team15} was utilized for producing bootstrap samples from the estimated principal component scores. These bootstrap samples are capable of mimicking the correlation within each univariate time series of principal component scores. Conditional on the estimated mean function and estimated functional principal components, the bootstrapped functional time series can be expressed as
\begin{align*}
\mathcal{X}_i^b(t)=\widehat{\mu}(t) + \sum^{n}_{k=1}\widehat{\beta}_{i,k}^b\widehat{\phi}_k(t),\qquad i=1,\dots,n,
\end{align*}
where $\big\{\widehat{\beta}_{1,k}^b,\dots,\widehat{\beta}_{n,k}^b\big\}$ represents the bootstrapped $k^{\text{th}}$ principal component scores, for $b=1,\dots,B$ where $B$ symbolizes the number of bootstrap replications. While this method is data-driven, considerations are also given to two residual bootstrap techniques described in the subsections~\ref{sec:far} and~\ref{sec:fkr}.

\subsection{Functional autoregressive (FAR) bootstrap}\label{sec:far}

When the stationarity condition satisfies for a functional time series $\left\{\mathcal{X}_1,\dots,\mathcal{X}_n\right\}$, a parametric or a nonparametric estimator of the conditional mean given by
\begin{align}
m(\mathcal{X}_{n-1},\dots,\mathcal{X}_{n-p}) = \text{E}(\mathcal{X}_n|\mathcal{X}_{n-1},\dots,\mathcal{X}_{n-p}), \label{eq:far_boot}
\end{align}
is able to capture correlation among functional time series. From a parametric bootstrapping aspect, \cite{Bosq00} proposed the functional autoregressive of order 1 (FAR(1)) and derived one-step-ahead forecasts that are based on regularized form of the Yule-Walker equations. Later, FAR(1) has been extended to FAR$(p)$, where the order $p$ can be determined via a hypothesis testing procedure of \cite{KR13}. \cite{KK16} proposed the functional moving average (FMA) process and introduced an innovations algorithm to obtain the best linear predictor. \cite{KKW16} extended the FAR and FMA processes to functional autoregressive moving average (FARMA). \cite{LRS16} considered long-range dependent functional time series and proposed a functional autoregressive integrated moving average process. Here, we consider the FAR model and sample with replacement from error functions,  where temporal dependence can be captured by functional autoregressive (FAR) of order $p$ (see e.g., \cite{Bosq00}). The FAR($p$) model can be expressed as:
\begin{equation*}
\mathcal{X}_{\omega} = \mu + \rho_1\left(\mathcal{X}_{\omega-1}-\mu\right) +\cdots+ \rho_p\left(\mathcal{X}_{\omega-p}-\mu\right)+\varepsilon_{\omega}, \end{equation*}
where $\varepsilon_{\omega}$ denotes an error term with mean zero; $(\rho_1,\dots,\rho_p)$ represent different lags autocorrelation operators; and $\omega=p+1,\dots,n$. Although \cite{KR13} presented a multistage testing procedure for determining the optimal order $p$ of a FAR process, it does not guarantee that $\varepsilon_{\omega}$ is completely serial uncorrelated. Instead, our bootstrap method uses the simplest FAR(1) model, where estimated error functions can then be resampled by the ME bootstrap, in order to capture any remaining temporal dependence exhibited in the error term. The FAR(1) model is given by
\begin{align}
\mathcal{X}_{\omega} = \mu + \rho \left(\mathcal{X}_{\omega-1} - \mu\right)+\varepsilon_{\omega}, \label{eq:far_2}
\end{align}
where $\rho=\gamma(1)/\gamma(0)$ represents the first-lag autocorrelation operator. Let $\mathcal{X}_{\omega}^c(t)$ be the centred functional data, then~\eqref{eq:far_2} can be conveniently re-expressed as 
\begin{equation*}
\mathcal{X}_{\omega}^c = \rho\mathcal{X}_{\omega-1}^c+\varepsilon_{\omega},
\end{equation*} 
where $^{c}$ symbolizes a centered functional time series.

Based on a set of centered functional time series, the variance and autocovariance functions $\gamma(0)$ and $\gamma(1)$ are estimated by the sample variance and autocovariance
\[
\widehat{\gamma}(0) = \frac{1}{n}\sum^n_{\omega=1}\mathcal{X}_{\omega}^c \otimes \mathcal{X}_{\omega}^c,\quad
\widehat{\gamma}(1) = \frac{1}{n}\sum^{n-1}_{\omega=1}\mathcal{X}_{\omega}^c\otimes \mathcal{X}_{\omega+1}^c,
\]
where $\otimes$ denotes a tensor product.

Assuming the FAR(1) is the correct parametric model, the best linear predictor of $\mathcal{X}_{\omega}^c$ given the past values $\mathcal{X}_{\omega-1}^c$ is given by 
\begin{equation*}
\mathcal{\widehat{X}}_{\omega}^c = \widehat{\rho} \mathcal{X}_{\omega-1}^c.
\end{equation*}
The one-step-ahead estimation error is then given by
\begin{equation*}
\widehat{e}_{\omega} = \mathcal{X}_{\omega}^c - \widehat{\mathcal{X}}_{\omega}^c, \qquad \omega = 2,\dots,n.
\end{equation*}
Similar to $\{\mathcal{X}_{2},\dots,\mathcal{X}_n\}$, $\{\widehat{e}_{2},\dots,\widehat{e}_n\}$ is a set of realizations of a stochastic process, where $\widehat{\widehat{e}}_{\omega} = \widehat{e}_{\omega} - \overline{\widehat{e}}_{\omega}$ be a centered stochastic process. Without such a centering, the resulting bootstrap approximation often has a random bias that does not vanish in the limit; see \cite{Lahiri03}. 

Since the FAR(1) model may not capture all dependence structure in the original functional time series, it is expected some residual dependence structure will be manifested in the centered residual functions. These centered residual functions can be bootstrapped through the ME bootstrap method given in Section~\ref{sec:me}. In the simulation study, we also examine data generating processes (DGPs) that differ from the FAR(1) model.

\subsection{Functional kernel regression (FKR) bootstrap}\label{sec:fkr}

Although the class of linear time series is quite rich, such as the FAR models, the class of nonlinear time series is vast. It is clear that the FAR bootstrap will generally not work well when applied to functional time series $\{\mathcal{X}_1,\dots,\mathcal{X}_n\}$ from a nonlinear time series. Under some smoothness assumption on conditional mean function $m(\cdot)$ in~\eqref{eq:far_boot}, this data-driven estimation can be performed under different nonparametric smoothing techniques, such as functional kernel smoothing (see, e.g., \cite{FV06}, \cite{Masry05}, \cite{PP00}). For some fixed order $p\in \mathbb{N}$, the error functions are given as
\begin{align}
\varepsilon_{\omega} = \mathcal{X}_{\omega} - m\left(\mathcal{X}_{\omega-1},\dots,\mathcal{X}_{\omega-p}\right), \label{eq:fkr}
\end{align}
where $\omega=p+1,\dots,n$, $m(\cdot)$ is an unknown smooth function, and $\varepsilon_{\omega}$ denotes error functions. The functional form of $m(\cdot)$ is often estimated in a data-driven manner. There are a growing amount of literature on the development of nonparametric functional estimators, such as functional Nadaraya-Watson estimator in \cite{FV06}, functional local linear estimator in \cite{BEM11}, functional $k$-nearest neighbor estimator in \cite{KV13} and distance-based local linear estimator in \cite{BDF10}. Throughout this paper, we illustrate the proposed method using the functional Nadaraya-Watson estimator because of its simplicity and mathematical elegance. 

Let $\mathcal{X}_{\omega}$ be the functional response, and $\bm{\mathcal{X}} = \big(\mathcal{X}_{\omega-1},\dots,$ $\mathcal{X}_{\omega-p}\big)$ be functional predictors. The functional Nadaraya-Watson estimator can be written as 
\begin{equation*}
\widehat{m}_n(x) = \frac{\sum^n_{\omega=p+1}K\left[\frac{d(x,\bm{\mathcal{X}})}{h}\right]\mathcal{X}_{\omega}}{\sum^n_{\omega=p+1}K\left[\frac{d(x,\bm{\mathcal{X}})}{h}\right]}, \qquad p\geq 1,
\end{equation*}
where $K(\cdot)$ is a symmetric real-valued kernel function defined on $R^+$ satisfying $\int_{-\infty}^{\infty}K(x)dx =1$;  $d(\cdot,\cdot)$ is a semi-metric used to measure distance between two functions and it has the properties $d(a,a) = 0$, but $d(a,b)=0 \centernot\implies a=b$ ; $h\in R^+$ represents a bandwidth associated with an infinite-dimensional function-valued predictor, and it controls the trade-off between squared bias and variance in the mean squared error. This estimator $\widehat{m}_n(x)$ is a weighted average of the observed response, where the weights are governed by a semi-metric distance and a bandwidth.

From~\eqref{eq:fkr}, the residual functions can be expressed as
\begin{equation*}
\widehat{e}_{\omega} = \mathcal{X}_{\omega} - \widehat{m}_n\left(\mathcal{X}_{\omega-1},\dots,\mathcal{X}_{\omega-p}\right), 
\end{equation*}
let $\widehat{\widehat{e}}_{\omega} = \widehat{e}_{\omega} - \overline{\widehat{e}}_{\omega}$ denotes the centered stochastic process. Similar to the FAR bootstrap, the FKR bootstrap also requires the optimal selection of order $p$. Following \cite{RAV15}, we also consider $p=1$ expressed as
\begin{align}
 \widehat{e}_{\omega} = \mathcal{X}_{\omega} - \widehat{m}_n\left(\mathcal{X}_{\omega-1}\right), \qquad \omega = 2,\dots,n.\label{eq:NFR}
\end{align}
Since~\eqref{eq:NFR} may not capture all dependence structure in the original functional time series, it is expected the residual dependence structure will be manifested in the centered residual functions. These centered residual functions can then be bootstrapped through the ME bootstrap method given in Section~\ref{sec:me}.

\section{Simulation study}\label{sec:4}

In Section~\ref{sec:2}, we introduce a kernel sandwich estimator for estimating long-run covariance of functional time series, used by the three bootstrap techniques described in Section~\ref{sec:3} to estimate the distribution of the long-run covariance of functional time series. In Section~\ref{sec:3.2}, we introduce a simulation setup, while the evaluation metrics are given in Section~\ref{sec:3.3}. Simulation results are presented in Section~\ref{sec:3.4}, where the accuracies of different resampling techniques are evaluated and compared. 

\subsection{Simulation DGPs}\label{sec:3.2}

In order to define the DGPs that we considered, let $\{B_i(t), -\infty<i<\infty, t\in [0,1]\}$ denote IID standard Brownian motions. Following an early work by \cite{RS16}, we generate functional time series according to
\begin{align*}
\text{FAR}_{\phi}(p)&: \mathcal{X}_i(t) = \sum^p_{j=1}\phi_j \mathcal{X}_{i-j}(t) + B_i(t),  \\
\text{FMA}_{\theta}(q)&: \mathcal{X}_i(t) = B_i(t) + \sum^q_{j=1}\theta_j B_{i-j}(t). 
\end{align*}
We consider two FAR processes, namely FAR$_{0.5}(1)$ and FAR$_{(-0.6, 0.09)}(2)$, and three functional moving average (FMA) processes, namely FMA$_{0.5}(1)$, FMA$_{0.5}(4)$ and FMA$_{0.5}(8)$.

\begin{figure}[!htbp]
\centering
{\includegraphics[width=8.5cm]{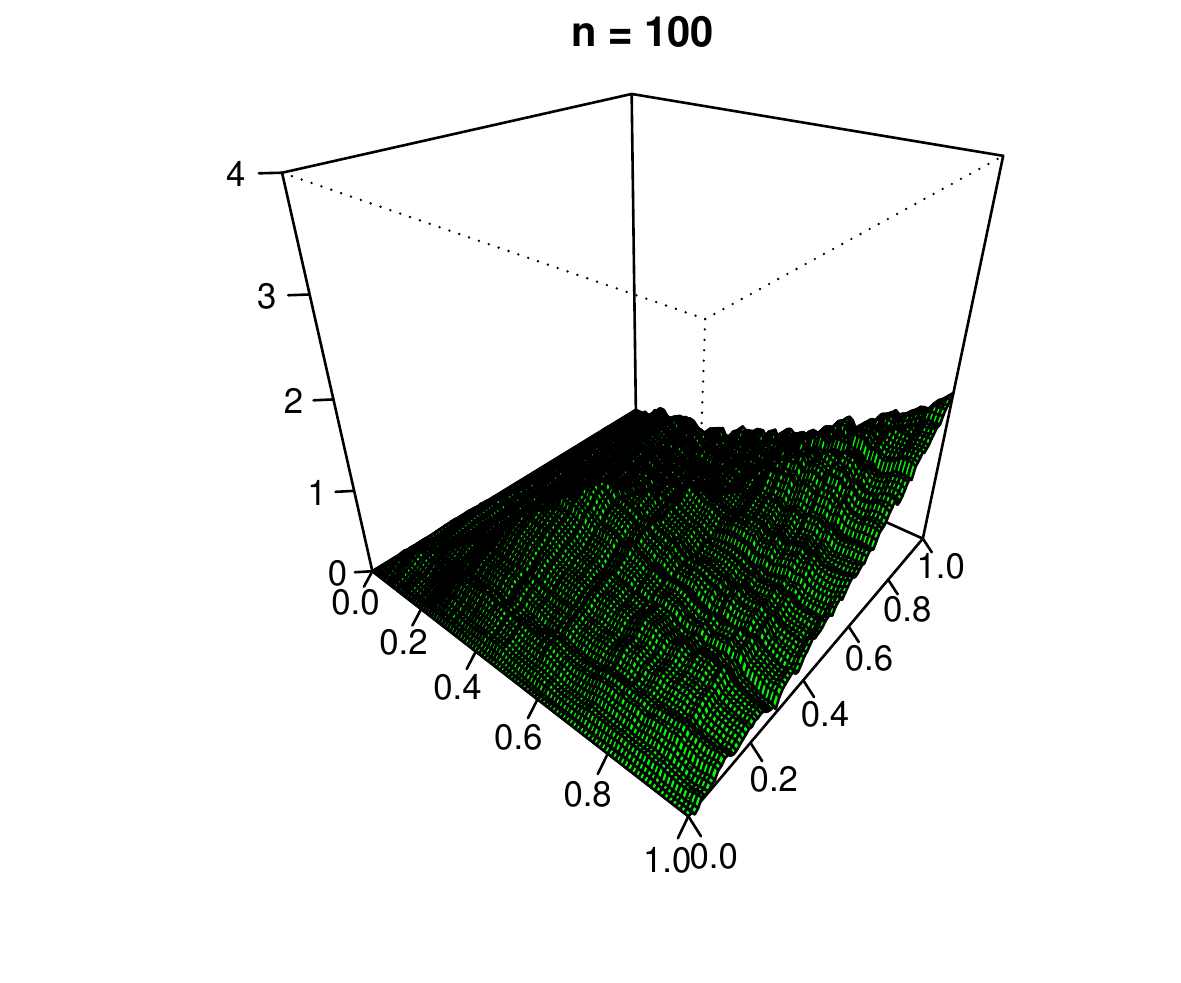}}
{\includegraphics[width=8.5cm]{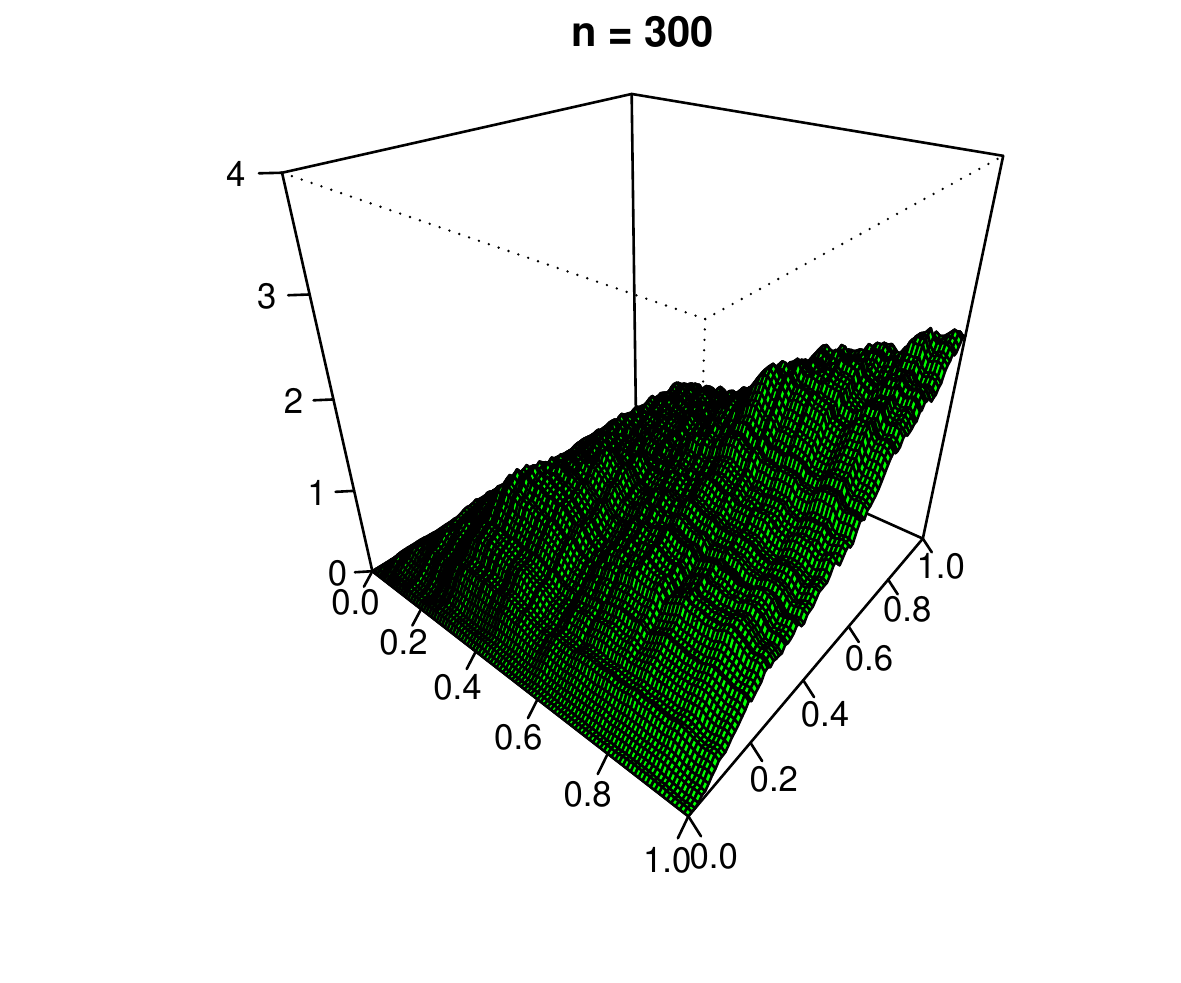}}
\\
{\includegraphics[width=8.5cm]{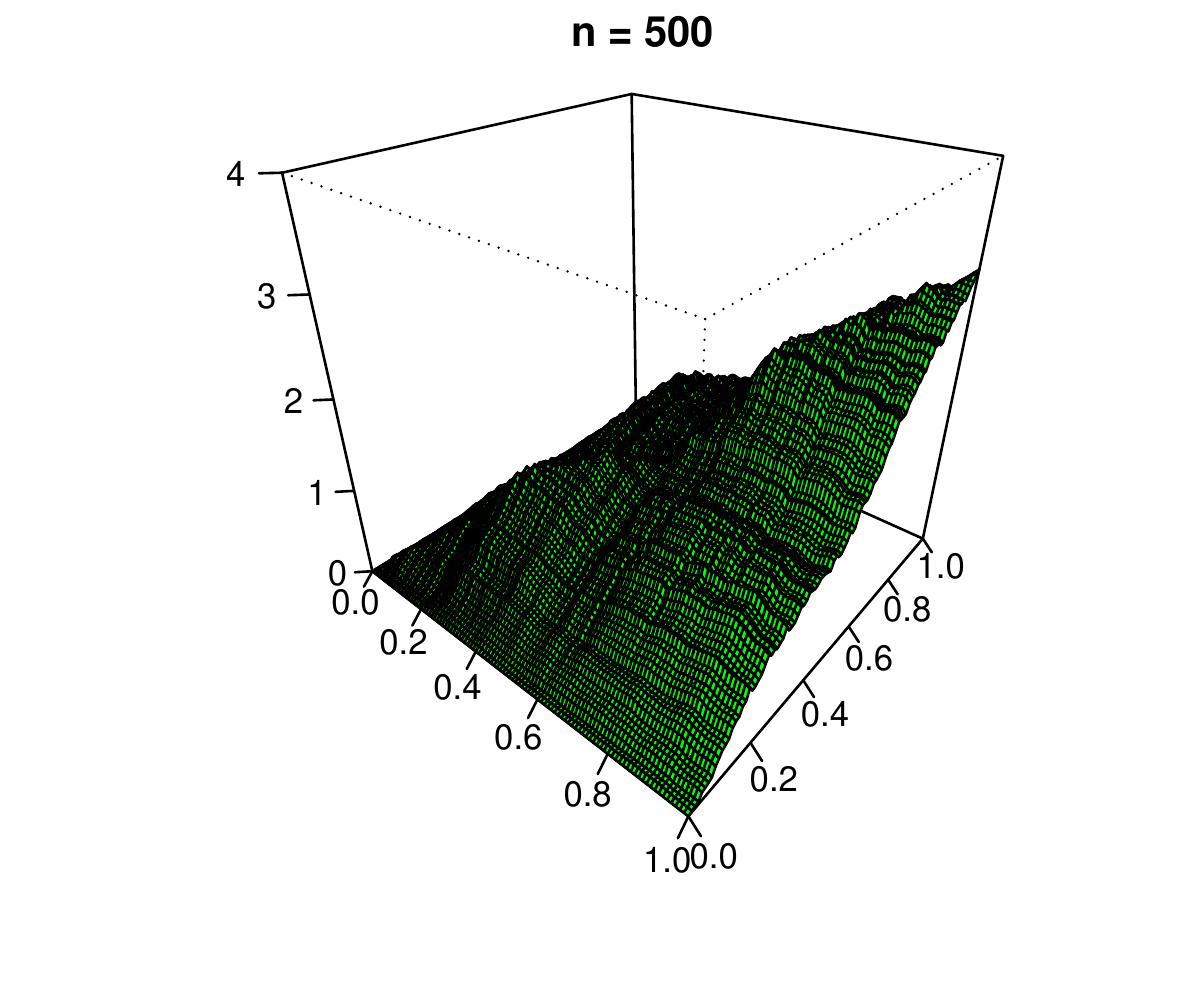}}
{\includegraphics[width=8.5cm]{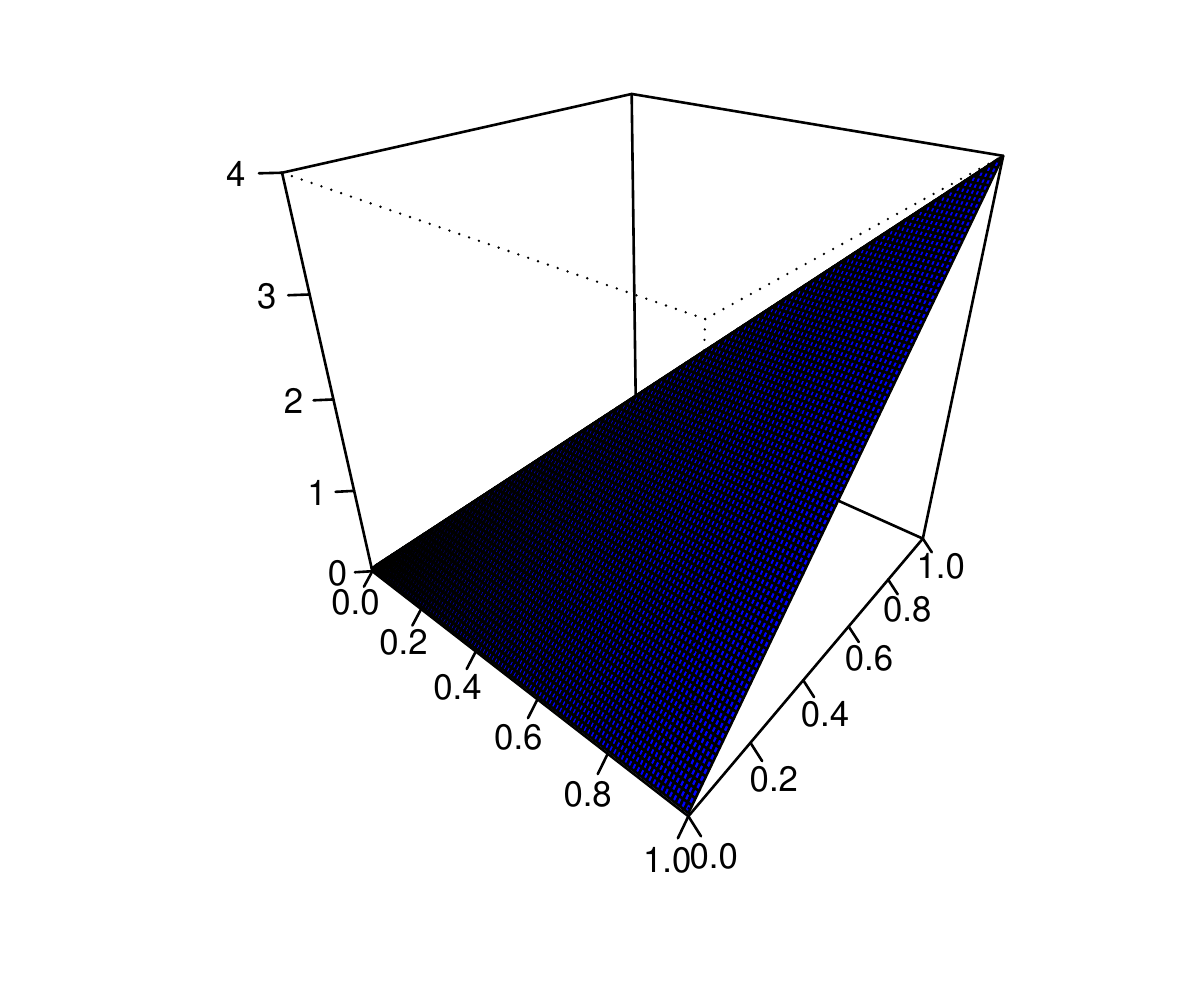}}
\caption{Lattice plots of the long-run covariance function estimates with the FAR$_{0.5}(1)$ simulated data, using the first-order Bartlett kernel with the proposed plug-in bandwidth for values of $n=100, 300$ and $500$ along with the theoretical long-run covariance (lower right).}\label{fig:1}
\end{figure}

As an illustration of the estimators $\widehat{C}_{h,q}$, Figure~\ref{fig:1} shows lattice plots of the sample long-run covariance estimates with the FAR$_{0.5}(1)$ simulated data, using the plug-in bandwidth selection procedure for $n=100, 300, 500$ and theoretical long-run covariance. 

\subsection{Simulation evaluation}\label{sec:3.3}

To evaluate the finite-sample performances of the three bootstrap methods, we investigate the distribution of the estimation errors between the sample estimated and theoretical long-run covariance functions, and between the sample estimated and bootstrapped sample long-run covariance functions. These can be expressed as
\begin{align*}
\widehat{D} &= \|\mathcal{K}(s,t) - \widehat{\mathcal{K}}(s,t)\|_2, \\
\widehat{D}^b &= \|\widehat{\mathcal{K}}(s,t) - \widehat{\mathcal{K}}^b(s,t)\|_2,
\end{align*}
where $\widehat{\mathcal{K}}^b(s,t)$ represents the bootstrapped sample long-run covariance function estimates, and $\|\epsilon(s,t)\|_2 = [\int \epsilon^2(s,t)dsdt]^{1/2}$. Given a sample $\{\mathcal{X}_1,$ $\mathcal{X}_2,\dots, \mathcal{X}_n\}$, we draw $R=200$ replications of bootstrapped samples using the three bootstrap methods, along with the IID bootstrap method of \cite{Shang15}; and the same pseudo-random seeds were used for all the methods in order to give the same simulation randomness. 

The $100(1-\alpha)\%$ bootstrap CIs of the long-run covariance function estimation error, is defined by calculating the cut-off value, $D_T$, such that the $100(1-\alpha)\%$ of the bootstrapped estimation errors are within a distance smaller than $D_T$. In our simulation study, the performance of bootstrap is evaluated through $R=200$ replications, the corresponding CIs are constructed based on $B=399$ repetitions for each replication. 

To calculate pointwise interval forecast accuracy, we utilize the interval score of \cite{GR07} (see also \cite{GK14}). A scoring rule for the pointwise interval forecast is given as
\begin{align*}
S_{\alpha}(\widehat{D}^l, \widehat{D}^u, D) = (\widehat{D}^u - \widehat{D}^l) & +  \frac{2}{\alpha}(\widehat{D}^l - D)I\{D < \widehat{D}^l\} \\
& + \frac{2}{\alpha}(D - \widehat{D}^u)I\{D > \widehat{D}^u\},
\end{align*}
where $I\{\cdot\}$ denotes a binary indicator function, and $\alpha$ denotes the levels of significance, such as $\alpha=0.05, 0.2, 0.5$ for their corresponding $95\%, 80\%, 50\%$ CIs, respectively. The interval score rewards a narrow CI, if and only if the true observation $D$ lies within the CI constructed from $\widehat{D}^b$. The optimal interval score is achieved when $D$ lies between $\widehat{D}^l$ and $\widehat{D}^u$ and the distance between $\widehat{D}^l$ and $\widehat{D}^u$ is minimal.

Averaged over $R=200$ replications, we compute the averaged interval score given by 
\begin{equation*}
\overline{S}_{\alpha} = \frac{1}{R}\sum_{r=1}^{R}S_{\alpha,r}. 
\end{equation*}
The best bootstrap method is the one that has the smallest averaged interval score.

\subsection{Simulation results}\label{sec:3.4}

In Table~\ref{tab:1}, we report the averaged interval scores obtained from four bootstrap methods for sample sizes $n=100, 200$, bootstrap repetitions $B=399$, and bootstrap replications $R=200$. 

\begin{table}[!htbp]
\centering
\tabcolsep 0.09in
\caption{Averaged interval scores based on $n=100, 200$, $B=399$ repetitions and $R=200$ replications. The smallest averaged interval scores are highlighted in bold.}\label{tab:1}
\begin{tabular}{@{}lrrrrrrrrr@{}}
\hline
 \multicolumn{5}{c}{$n=100$} & \multicolumn{5}{c}{$n=200$} \\
 $\alpha$ & \multicolumn{4}{c}{Bootstrap method} & & \multicolumn{4}{c}{Bootstrap method} \\
 DGP  & IID & ME & FAR & FKR & IID & ME & FAR & FKR  \\\hline
  $\underline{\alpha=0.05}$ & & & & &  & & & & \\
  FAR$_{0.5}$(1) & 6.1745 & 0.1401 & \textBF{0.0758} & 0.0977 &  14.0217 & 0.1207 & \textBF{0.0642} & 0.0764 \\ 
  FAR$_{(-0.6,0.09)}$(2) & 6.8091 & 0.1474 & \textBF{0.0688} & 0.0941 & 14.7645 & 0.1214 & \textBF{0.0568} & 0.0703 \\
  FMA$_{0.5}$(1) & 0.0958 & 0.0072 & \textBF{0.0042} & 0.0051 & 0.1689 & 0.0061 & \textBF{0.0033} & 0.0038 \\ 
  FMA$_{0.5}$(4) & 65.3362 & 0.3658 & \textBF{0.0842} & 0.1163  & 94.9467 & 0.3185 & \textBF{0.0693} & 0.0886 \\ 
  FMA$_{0.5}$(8) & 819.4318 & 4.2373 & \textBF{0.4305} & 0.6141 & 1250.9981 & 4.2912 & \textBF{0.3914} & 0.5441 \\ \\
  $\underline{\alpha=0.2}$ & & & & \\
  FAR$_{0.5}$(1) & 2.2643 & 0.1709 & \textBF{0.0997} & 0.1186 & 4.1063 & 0.1336 & \textBF{0.0762} & 0.0872 \\  
  FAR$_{(-0.6,0.09)}$(2) & 2.4869 & 0.1725 & \textBF{0.0923} & 0.1138 & 4.3445 & 0.1309 & \textBF{0.0678} & 0.0792 \\
  FMA$_{0.5}$(1) & 0.0649 & 0.0080 & \textBF{0.0050} & 0.0059 & 0.0772 & 0.0061 & \textBF{0.0036} & 0.0041 \\ 
  FMA$_{0.5}$(4) & 17.6960 & 0.4312 & \textBF{0.1155} & 0.1463 & 24.8471 & 0.3463 & \textBF{0.0860} & 0.1066 \\ 
  FMA$_{0.5}$(8) & 213.7299 & 5.0534 & \textBF{0.5839} & 0.7973 & 320.1938 & 4.7279 & \textBF{0.5255} & 0.6861 \\ 
\\
  $\underline{\alpha=0.5}$ & & & & \\
  FAR$_{0.5}$(1) & 1.3222 & 0.1671 & \textBF{0.0996} & 0.1168 & 1.9924 & 0.1303 & \textBF{0.0753} & 0.0858 \\ 
  FAR$_{(-0.6,0.09)}$(2) & 1.4446 & 0.1659 & \textBF{0.0923} & 0.1109 & 2.1165 & 0.1258 & \textBF{0.0670} & 0.0776 \\
  FMA$_{0.5}$(1) & 0.0467 & 0.0074 & \textBF{0.0048} & 0.0056 & 0.0495 & 0.0055 & \textBF{0.0034} & 0.0039 \\ 
  FMA$_{0.5}$(4) & 7.9032 & 0.4273 & \textBF{0.1171} & 0.1462 & 10.6051 & 0.3373 & \textBF{0.0862} & 0.1066 \\ 
  FMA$_{0.5}$(8) & 90.9713 & 5.0452 & \textBF{0.6252} & 0.8171 & 132.6385 & 4.7377 & \textBF{0.5480} & 0.6977 \\ 
\hline
  \end{tabular}
\end{table}

Subject to the same random seed, the FAR bootstrap method performs the best with the smallest averaged interval score, followed closely by the FKR bootstrap method. For comparison, we also include the IID bootstrap method of \cite{Shang15}, where each set of the principal component scores is randomly sampled with replacement. It produces rather unsatisfactory results with the largest averaged interval scores in all DGPs. This demonstrates that any naive application assuming that the functional data are IID when in fact they are not, can be disastrous. In contrast, the proposed three bootstrap methods were implemented by preserving the data structure and show empirically bootstrap consistency. As the sample size $n$ increases, the averaged interval scores become smaller. 

When the DGPs were simulated from the FAR(1) models, the FAR bootstrap method performs the best, since the assumed model is correct. To our surprise, the FAR bootstrap method is also favorable when the DGPs were simulated from the FAR$_{(-0.6, 0.09)}(2)$ and three FMA models. In the former case, an explanation is that the DGP is linear in structure, thus a functional autoregression model even with a wrongly chosen order would be preferable than a functional kernel regression. In the latter case, an explanation is that stationary and invertible FMA models can be re-expressed in terms of a FAR model. As the temporal dependence increases from FMA(1) to FMA(4) to FMA(8), the averaged interval scores increase for all bootstrap methods. It reveals the difficulty of all the bootstrap methods considered for modeling long-range dependent functional time series.

\section{Intraday PM$_{10}$ curves}\label{sec:5}

As a vehicle of illustration, intraday PM$_{10}$ concentrations were considered. The observations are half-hourly measurements of concentration of PM with an aerodynamic diameter of less than 10um, in ambient air taken in Graz, Austria from 1/October/2010 until 31/March/2011 (see also \cite{HKH15}). Based on this observed time period, we convert $N=8,736$ discrete univariate time series points into $n=182$ daily curves. 

\begin{figure}[!htbp]
\centering
\subfloat[A univariate time series display]
{\includegraphics[width=8.7cm]{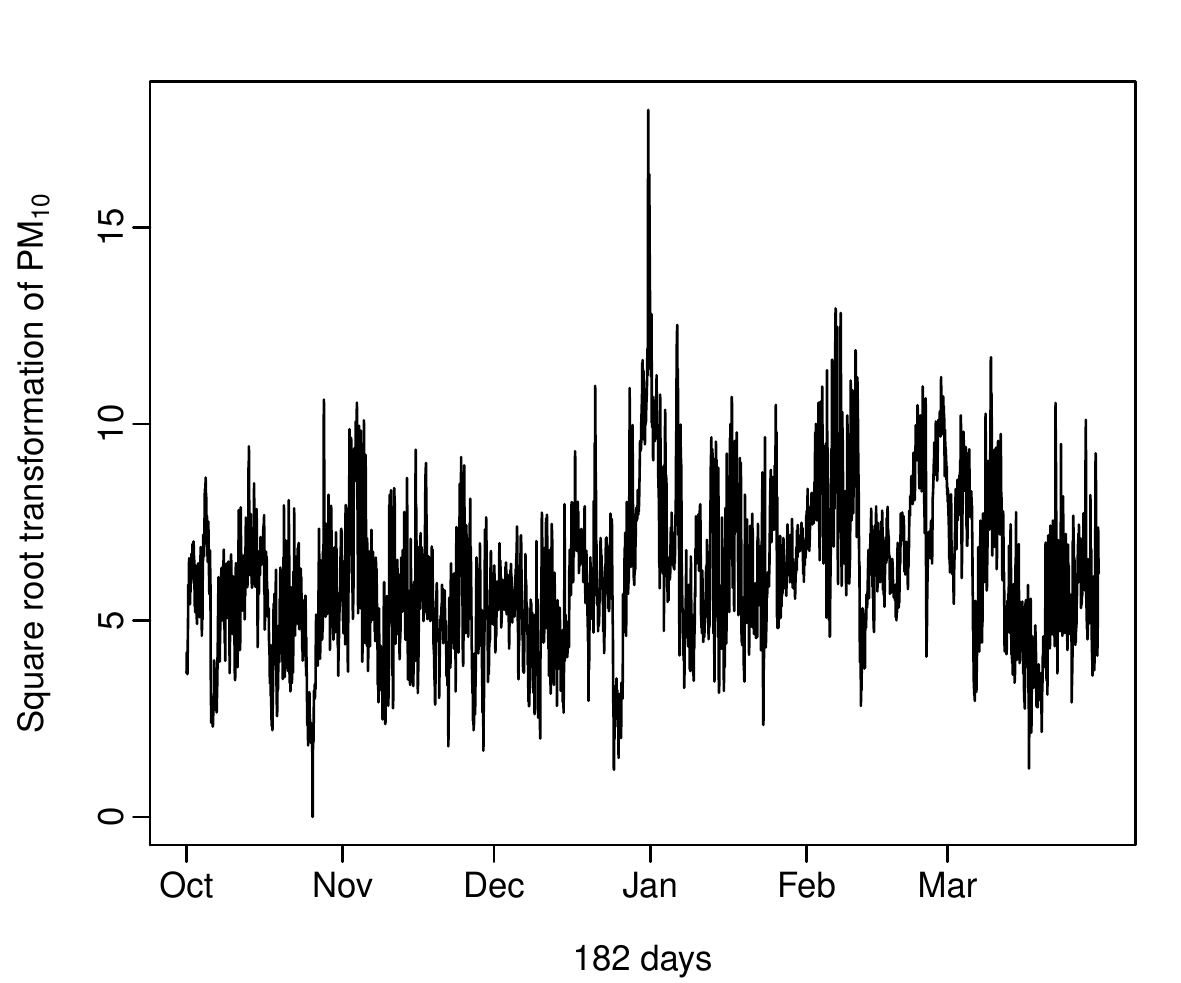}\label{fig:uni}}
\\
\subfloat[A functional time series display]
{\includegraphics[width=8.7cm]{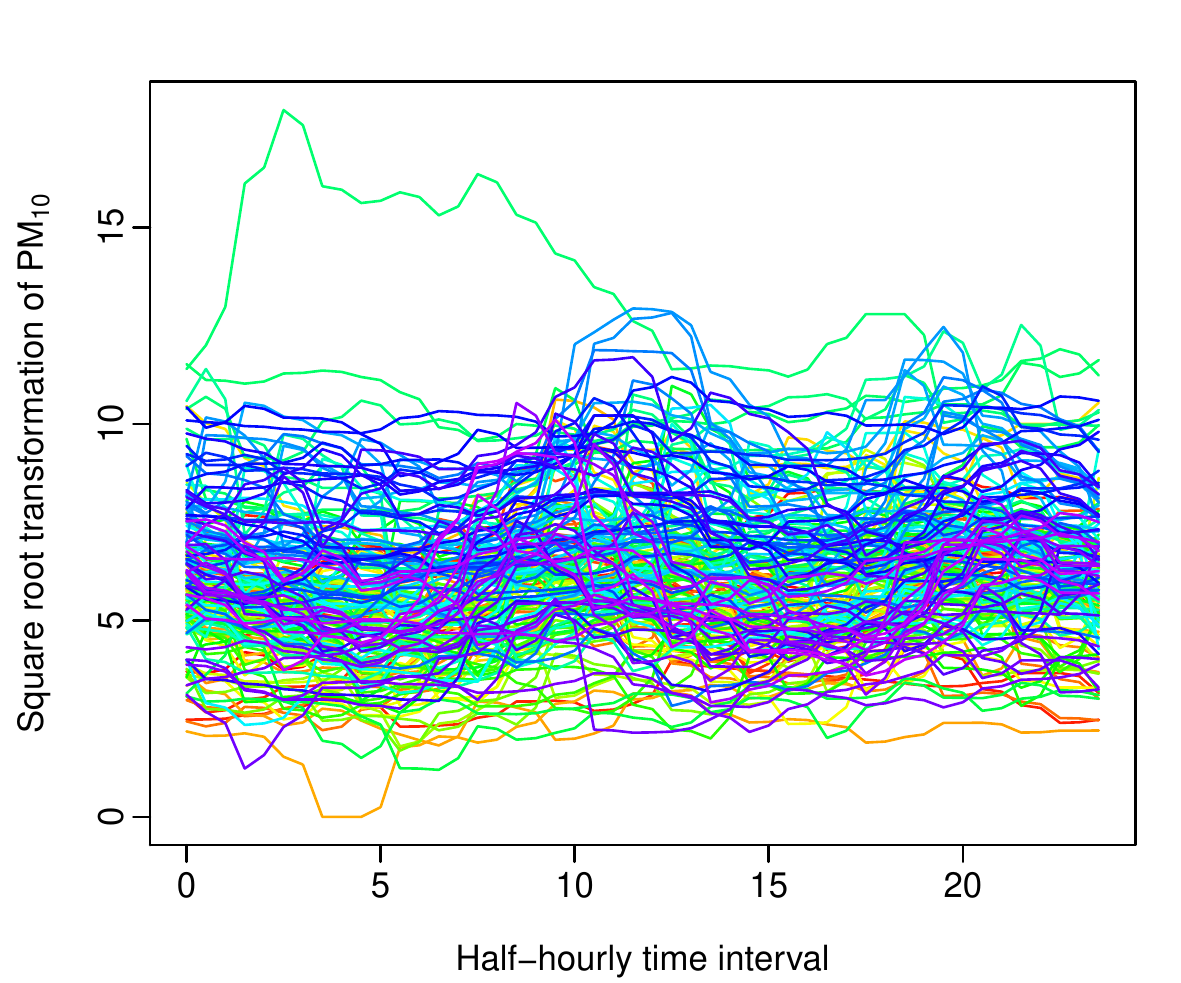}\label{fig:fts}}
\caption{Graphical displays of intraday measurements of the PM$_{10}$ from 1/October/2010 to 31/March/2011.}\label{fig:UK_fert}
\end{figure}

Let $\{Z_{w}, w\in [1,N]\}$ be a seasonal univariate time series, which has been observed at $N$ equispaced time points. When the seasonal pattern is strong, one way to model the time series nonparametrically is to use ideas from functional data analysis (see also \cite{FGV02}). We divide the observed time series into $n$ trajectories, and then consider each trajectory of length $p$ as a curve rather than as $p$ distinct points. The functional time series is given by
\begin{equation*}
\mathcal{X}_i(t_j) = \{Z_w, \ w\in (p(i-1), pi]\}, \qquad i = 1,\dots,n,
\end{equation*}
where $j=1\dots,48$ denotes $p=48$ discrete points, and $0<t_1\leq t_2\leq \cdots \leq t_{48}=24$. A univariate time series display of intraday pollution curves is given in Figure~\ref{fig:uni}, with the same data shown in Figure~\ref{fig:fts} as a time series of functions.

Using the stationarity test proposed by \cite{HKR14}, we carried out a hypothesis testing procedure with the null hypothesis that the functional time series is stationary. Since the $p$-value = $0.078>0.05$, we conclude that this functional time series is stationary. 

\begin{figure}[!htbp]
\centering
\subfloat[Sample estimated covariance]
{\includegraphics[width=8.4cm]{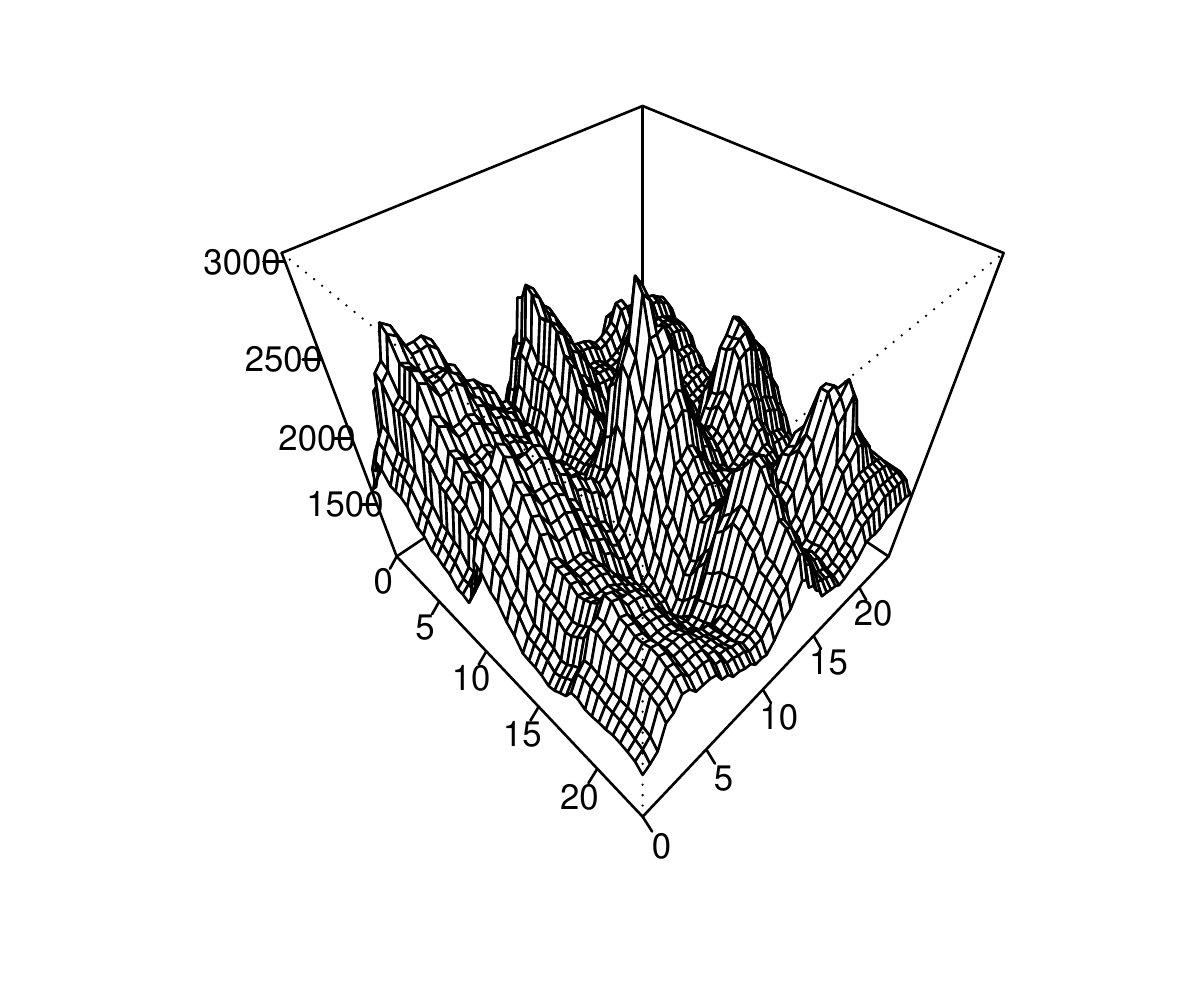}\label{fig:3a}}
\quad
\subfloat[Estimated lower bound]
{\includegraphics[width=8.4cm]{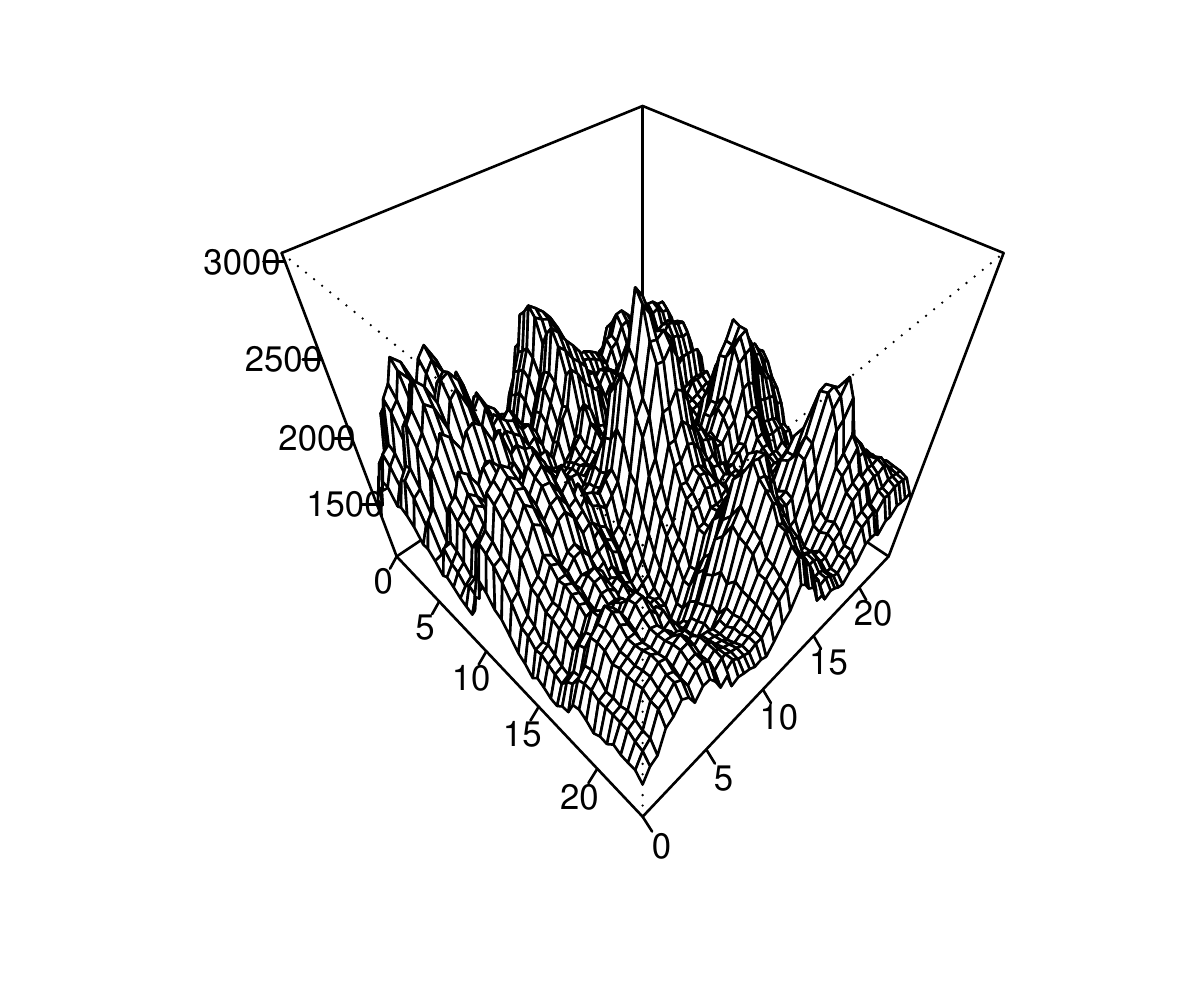}\label{fig:3b}}
\quad
\subfloat[Estimated upper bound]
{\includegraphics[width=8.4cm]{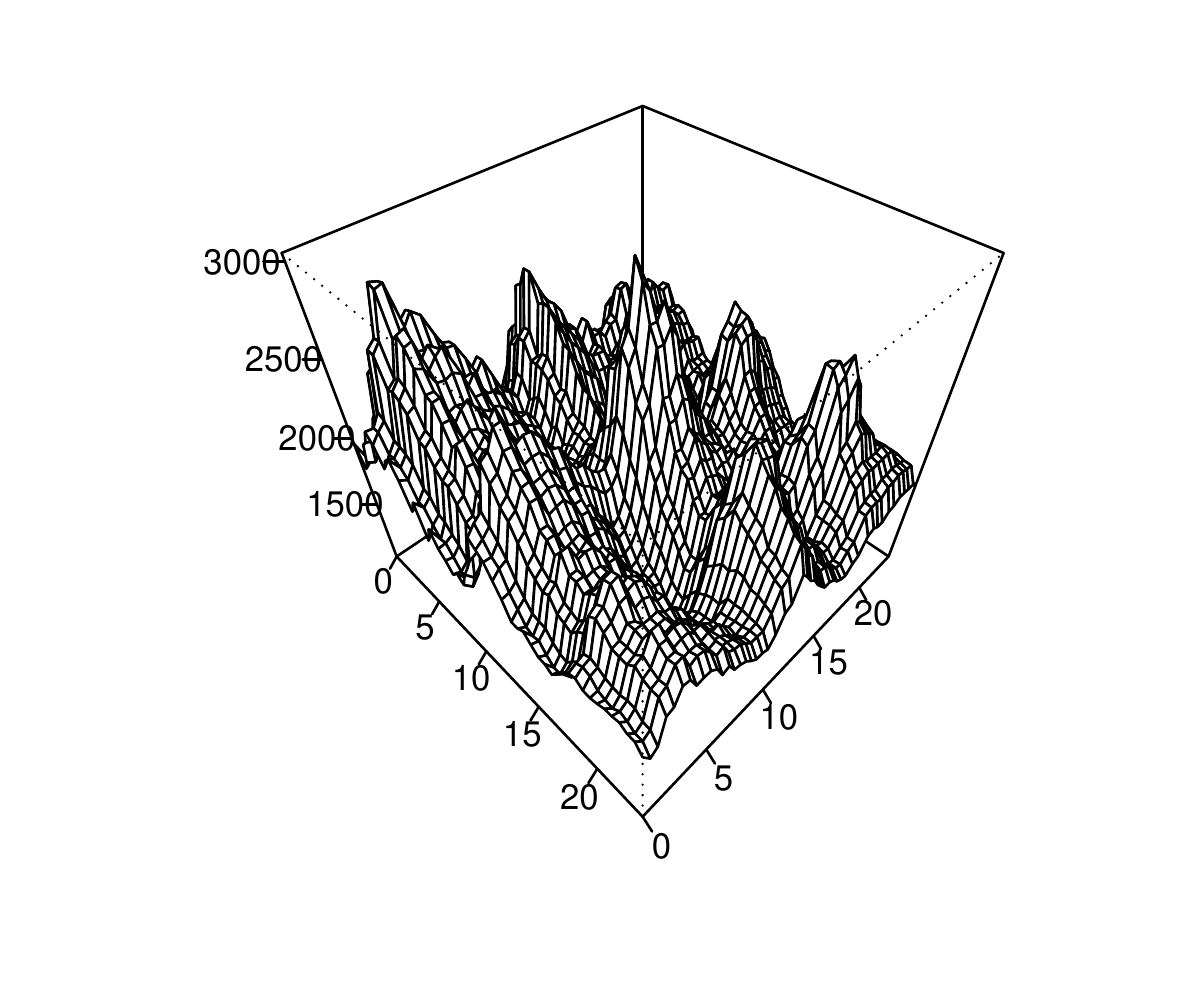}\label{fig:3c}}
\caption{Sample long-run covariance estimated by the kernel sandwich estimator, and the estimated 80\% lower and upper bounds of the estimated long-run covariance using Bartlett kernel and the plug-in bandwidth.}\label{fig:ausfert}
\end{figure}

Our aim is to apply a bootstrap method to construct the 80\% CI of the estimated long-run covariance of the PM$_{10}$ intraday data set. Among the bootstrap methods considered, we implemented the FAR bootstrap method, as it has the best finite-sample performance in our simulation study. While the sample estimated long-run covariance is shown in Figure~\ref{fig:3a}, the lower and upper bounds of the 80\% CIs are shown in Figures~\ref{fig:3b} and~\ref{fig:3c}, respectively. 

\section{Conclusion}\label{sec:6}

We present three bootstrap methods to visualize the distribution of a descriptive statistic of functional time series. Since functional time series are intrinsically infinite dimension, a functional principal component analysis was used to reduce dimensionality and the autocorrelation in the original functional time series is manifested in the principal component scores. By bootstrapping these scores via the ME, bootstrapped functional observations that preserve the temporal dependence were obtained, conditional on the estimated mean function and estimated functional principal components. While this method is purely data-driven, we also consider the first-order FAR and FKR to first capture simple linear and non-linear temporal dependence structures, before bootstrapping centered residual functions by the ME. 

Through some Monte-Carlo simulation studies, we illustrate the improved averaged interval scores as the sample size increases from $n=100$ to $200$ for all proposed bootstrap methods. In the presence of autocorrelation, it is advantageous to consider bootstrap methods that are appropriate for functional time series, as they would result in much smaller estimation errors and hence better estimation accuracy of a population statistic, such as the long-run covariance considered here. 

One limitation of the simulation study is that the DGPs are restricted to a linear functional form, which is common in \cite{Bosq00} and \cite{HRW16}. However, non-linear functional time series analysis is an important research topic that deserves future investigation, but it is currently in its infancy. The other limitation of the proposed bootstrap methods is that we can not establish the bootstrap consistency. It is a topic that we aim to address in future research.

Illustrated by the intraday PM$_{10}$ data set in Graz, it is shown that the first-order FAR bootstrap procedure provides an effective descriptive tool to the distributional analysis of the estimated long-run covariance of a stationary functional time series. It is our hope that the bootstrap methods, including the ones proposed here, will receive increasing popularity in functional time series analysis, where the object of interest is on the distribution of functional estimators.

A future research direction is to consider high-order bootstrapping for functional time series. Recent advances in statistical and econometric theories show that iterating the bootstrap principle brings further refinements upon the single bootstrapping (see, for example, \cite{DM07}). Iterating the bootstrap principle reduces the dependence structure between the probability distribution of the resamples and the unknown DGP. For instance, double bootstrap has typically higher order accuracy than single bootstrap, but at a much higher computational cost. Addressing this computational challenge and extending high-order bootstrapping to functional time series remain a future research.

\bibliographystyle{agsm}
\bibliography{master}

\end{document}